\documentclass[11pt]{article}

\usepackage{amssymb,amsmath,graphicx}

\begin{document}

\begin{titlepage}

\begin{center}

\vspace{0.8cm}

{\Large Affleck-Dine leptogenesis via multiscalar evolution\\
in a supersymmetric seesaw model}

\vspace{0.8cm}

\vspace{1.3cm}

{\bf Masato Senami}\footnote{senami@icrr.u-tokyo.ac.jp.
Now Department of Micro Engineering, Kyoto University. 
}
and
{\bf Tsutomu Takayama}\footnote{tstkym@icrr.u-tokyo.ac.jp}
 \\

\vspace{1cm}

{\it
ICRR, University of Tokyo, Kashiwa 277-8582, Japan
}

\vspace{1.5cm}

\abstract{
A leptogenesis scenario in a supersymmetric standard model extended with
introducing right-handed neutrinos is reconsidered.
Lepton asymmetry is produced in the condensate of a right-handed sneutrino
via the Affleck-Dine mechanism.
The $LH_u$ direction develops large value
due to a negative effective mass induced by the right-handed sneutrino condensate
through the Yukawa coupling of the right-handed neutrino,
even if the minimum during the inflation is fixed at the origin.
The lepton asymmetry is nonperturbatively transfered to
the $LH_u$ direction by this Yukawa coupling.
}

\end{center}
\end{titlepage}

\setcounter{footnote}{0}

\section{Introduction}

The origin of the baryon asymmetry of the universe is one of unsolved problems.
The existence of the baryon asymmetry is confirmed in several ways.
Among them, the observation of the cosmic microwave background constrains
the amount of the baryon asymmetry with considerable accuracy.
It is given in terms of the baryon-to-entropy ratio as \cite{Spergel:2006hy}
\begin{eqnarray}
	\label{eq:constraint_baryon-to-entropy_ratio}
	\frac{n_B}{s}= (8.7{\pm 0.3}) \times10^{-11}
\end{eqnarray}
where $s$ is the entropy density of the universe.

The origin of neutrino masses is another problem of the standard model (SM).
By observation of neutrino oscillations \cite{Experiments, Fogli:2005cq},
it is confirmed that at least two flavors of neutrinos have non-zero masses.
On the other hand, the sum of masses of three flavors of neutrinos is constrained as 
$\sum m_\nu < 2\,{\mathrm{eV}}$ by cosmological observations
\cite{Spergel:2006hy,Ichikawa:2004zi}.
Hence, the SM should be extended to explain
nonzero neutrino masses and the smallness of them.
Introducing heavy right-handed Majorana neutrinos provides a good explanation
for the problem of neutrino masses via the seesaw mechanism \cite{seesaw}.

Heavy right-handed Majorana neutrinos also provide an attractive solution 
of the origin of the baryon asymmetry by leptogenesis \cite{Fukugita:1986hr}.
In the leptogenesis scenario, lepton asymmetry is generated at first,
and then the sphaleron process partially transfers it into the baryon asymmetry.
In the thermal leptogenesis scenario,
the lepton asymmetry is generated via the out-of-equilibrium decay
of heavy right-handed Majorana neutrinos produced in the primordial thermal bath
\cite{Fukugita:1986hr,leptogenesis}.
Therefore, this scenario requires the reheating temperature $T_R$ after the inflation
to be higher than the mass of the lightest right-handed neutrino.
However, $T_R \lesssim 10^8 \,{\mathrm{GeV}}$ is required
to avoid the overproduction of the gravitino
in supersymmetric (SUSY) theories \cite{gravitino}.
The production of the sufficient lepton asymmetry is very difficult in SUSY models
due to this constraint \cite{leptogenesis}.

Hence, several alternative leptogenesis scenarios have been considered in SUSY theories.
The Affleck-Dine mechanism \cite{Affleck:1984fy,Dine:1995kz,Enqvist:2003gh},
which produces asymmetry between particle and antiparticle
in the condensate of a scalar field, is one of interesting scenarios,
since many species of scalar particles exist in SUSY models.
In the Affleck-Dine leptogenesis along the $LH_u$ flat direction~\cite{Murayama:1993em},
the source of lepton number violation originates in the operator of the neutrino mass.
However, in the SUSY seesaw model,
if the mass of a right-handed neutrino is smaller than
the Hubble parameter at the end of the inflation $H_{\rm inf}$,
the $LH_u$ direction is not flat due to the $F$-term potential
from the Yukawa coupling of the right-handed neutrino\footnote{
If a neutrino Yukawa coupling is very tiny, e.g. in the Dirac neutrino model,
the $LH_u$ direction is approximately flat
and can have large value during inflation.
The Affleck-Dine leptogenesis in the Dirac neutrino model 
can explain the baryon asymmetry without the lepton number violation
except for the sphaleron process \cite{Senami:2007up}.}.
The right-handed neutrino mass should be less than about $10^{13}$\,GeV
for $m_\nu \sim 1$\,eV if the Yukawa coupling is less than the unity.
For the chaotic inflation, which predicts $H_{\rm inf} \sim 10^{13}$\,GeV,
it is severe that the right-handed neutrino masses dominate over $H_{\rm inf}$.

In addition to right-handed neutrinos,
there are right-handed sneutrinos $\tilde N$ in the SUSY seesaw model.
The potential for $\tilde N$ is flat except for SUSY mass terms,
if $R$-parity is conserved.
Therefore, quantum fluctuation of a right-handed sneutrino can be large
at the end of the inflation \cite{Murayama:1993em},
as long as the right-handed neutrino mass is smaller than $ H_{\rm inf}$.
Hence, the right-handed sneutrino can have
the large number density at the end of the inflation.
Thus, if $CP$-violation in the $\tilde N$ decay
into leptons and anti-leptons is large enough,
sufficient lepton asymmetry can be produced.

Later, this scenario was drastically changed after
SUSY breaking in the early universe was reported \cite{Dine:1995uk}.
Due to the negative Hubble mass square of a right-handed sneutrino,
minima of the potential of the right-handed sneutrino
is largely deviated from the origin during the inflation.
Therefore, the right-handed sneutrino has large value.
After the inflation, this scalar field condensate begins coherent oscillations.
If there exists $CP$-violation in the potential, e.g. $B$-term of $\tilde N$,
this condensate acquires the particle number asymmetry for the right-handed sneutrino
via the Affleck-Dine mechanism.
In Ref.\,\cite{Senami:2002jg}, one scenario along with multidimensional
Affleck-Dine mechanism \cite{Senami:2002kn} is reported.
The asymmetry of $\tilde N$ is nonperturbatively transferred to the $LH_u$ direction,
if the $LH_u$ direction is approximately flat,
that is, both $\tilde N$ and $LH_u$ directions have large value during the inflation.
In this scenario, the lepton asymmetry can be generated
without $CP$-violation in the right-handed sneutrino decay.

On the other hand, in Ref.\,\cite{Allahverdi:2004ix}, Allahverdi and Drees reported 
that the particle number asymmetry of $\tilde N$ can be transfered
by perturbative decay of $\tilde N$
even if the $LH_u$ direction is not approximately flat
or has a positive Hubble mass square, i.e.,
without the evolution of the scalar field along the $LH_u$ direction.
Because of the Majorana nature of the right-handed neutrino,
right-handed sneutrinos decay with generating
$\pm1$ lepton number in the same decay rate if SUSY is conserved.
These two decay rates are deviated from each other
because of SUSY breaking by thermal effects in the early universe.
Therefore, non-zero lepton number can be generated
in compensation for some parameter tuning.

However, we found in this work that the evolution of the $LH_u$ direction
is induced by a negative effective mass
given by the right-handed sneutrino condensate,
even if the $LH_u$ direction is not approximately flat
or has a positive Hubble mass square.
Therefore, the evolution of the $LH_u$ direction cannot be neglected
in broad parameter region.
Thus, the evolution of the scalar fields and the lepton asymmetry is complicated,
like the scenario in Ref.\,\cite{Senami:2002jg}.
In this paper, we reconsider the scenario discussed in Ref.\,\cite{Allahverdi:2004ix},
including the evolution of the $LH_u$ direction.
We will see that the lepton asymmetry in the right-handed sneutrino condensate 
can be nonperturbatively transfered to the $LH_u$ direction condensate
via the interaction between these scalar fields.
We also summarize the condition that
the evolution of the $LH_u$ direction is relevant.

The rest of this paper is organized as follows.
In the next section, we summarize the set-up of this scenario.
In Section 3, we discuss the evolution of scalar fields.
In Section 4, we consider the evolution of the lepton asymmetry.
The resultant baryon asymmetry is discussed in Section 5.
We also comment on the difference between this scenario and the previous one.
Finally, we summarize the result in Section 6.

\section{Model}

We consider the SUSY seesaw model.
The superpotential is given by 
\begin{eqnarray}
	\label{eq:superpotential}
	W = W_{\mathrm{MSSM}} + y_\nu NLH_u + \frac{M_N}{2}NN + 
	\frac{\lambda}{4M_{\mathrm{Pl}}}NNNN,
\end{eqnarray}
where $W_{\mathrm{MSSM}}$ is the superpotential of the Minimal SUSY SM (MSSM),
$y_\nu$ is the coupling between right-handed neutrinos and left-handed leptons,
and $M_N$ is the mass matrix of right-handed neutrinos.
We included non-renormalizable superpotential of $N$ with coupling constant given by $\lambda$,
which is important for successful baryogenesis in this scenario.
For simplicity, only one flavor of right-handed sneutrino is considered.
This can be naturally realized by assuming that three neutrino masses satisfies
the condition, $M_N = M_{N_1} < H_{\mathrm{inf}} < M_{N_2} < M_{N_3}$.
Hence, we consider only one flavor of left-handed leptons coupling 
to the right-handed sneutrino. 
We choose $y_\nu$ and $M_N$ to be real and positive by redefining superfields.
We assign lepton number $-1$ to $\tilde{N}$ although
right-handed neutrinos cannot carry $U(1)$ charge because of their Majorana nature.
The lepton number of $\tilde{N}$ is violated by $B$-terms and non-renormalizable terms,
but this violation is suppressed by SUSY breaking effects or cut-off scale $M_{\mathrm{Pl}}$.
Therefore, we can assign the lepton number to $\tilde N$ in this work.

A light neutrino mass is given by the seesaw mechanism as,
\begin{eqnarray}
	\label{eq:seesaw}
	m_\nu = \frac{y_\nu^2 v^2}{M_N}
	\sim 10^{-2} \, {\mathrm{eV}} \left( \frac{y_\nu}{10^{-2}} \right)^2 
	\left( \frac{10^{11}{\mathrm{GeV}}}{M_N} \right).
\end{eqnarray}
Here $v\sim{\mathcal{O}}(100)\,{\rm GeV}$ is the vacuum expectation value (vev) of $H_u$.

We consider the evolution of the right-handed sneutrino $\tilde{N}$ 
and the $LH_u$ direction parametrized by a complex scalar field $\phi$,
namely,
\begin{eqnarray}
	\label{eq:flat-direction}
	\tilde L =\frac{1}{\sqrt{2}}\left(
	\begin{array}{c}	
		\phi \\ 0
	\end{array}
	\right),
	H_u=\frac{1}{\sqrt{2}}\left(
	\begin{array}{c}
		0 \\ \phi
	\end{array}
	\right).
\end{eqnarray}
The $LH_u$ direction is $D$-flat,
but not $F$-flat due to the Yukawa coupling of the neutrinos.
Nevertheless, the evolution of this direction gives
an important effect on the leptogenesis.
Hence, in this work we take this direction into consideration.

Including Hubble-induced SUSY-breaking effects and thermal-corrections,
the scalar potential is given by
\begin{eqnarray}
	\label{eq:potential}
	V(\phi,\tilde{N}) &=& \frac{y_\nu^2}{4}|\phi|^4 + M_N^2 |\tilde{N}|^2
	+ y_\nu^2 |\phi|^2 |\tilde{N}|^2 + \frac{\lambda^2}{M_{\mathrm{Pl}}^2}|\tilde{N}|^6  \nonumber \\
	&& + \left[ \left( \frac{y_\nu}{2} M_N {\phi}^2\tilde{N}^* 
	+ \frac{y_\nu\lambda}{2M_{\mathrm{Pl}}}\phi^2\tilde{N}^{*3}
	+ \frac{\lambda M_N}{M_{\mathrm{Pl}}} \tilde{N}{\tilde{N}^{*3}} \right) + h.c. \right] \nonumber \\
	&& \nonumber \\
	&&+ c_\phi H^2 |\phi|^2 - c_NH^2 |\tilde{N}|^2 \nonumber \\
	&& \nonumber \\
	&&+ \left[ \left( \frac{b H}{2}M_N \tilde{N}^2 
	+ \frac{a_y y_\nu}{2}H \phi^2\tilde{N} + \frac{a_\lambda \lambda}{4M_{\mathrm{Pl}}}H\tilde{N}^4\right) 
	+ h.c. \right] \nonumber \\
	&& \nonumber \\
	&&+ V_{\mathrm{th}}(\phi) ,
\end{eqnarray}
where $H$ is the Hubble parameter and we ignored low-energy SUSY-breaking terms,
which are not relevant to this leptogenesis.
The first and second lines are the $F$-term potential from the superpotential (\ref{eq:superpotential}),
while the third and fourth lines are Hubble-induced SUSY breaking terms.
Here we assumed that the Hubble-induced mass of $\phi$ is positive 
in order to set it at the origin during the inflation even for small $y_\nu$,
while that of $\tilde{N}$ is negative in order to give large value.
Therefore, real parameters $c_\phi \sim 1$ and $c_N \sim 1$ are both assumed to be positive.
Complex parameters $a_y$, $a_\lambda$ and $b$ determine
magnitude of the Hubble-induced $A$-terms and $B$-term, respectively.
Absolute values of these constants are typically ${\mathcal{O}}(1)$ during the inflation.
However, since these terms are generally indued by coupling with the inflaton,
they oscillate rapidly by the oscillation of the inflaton after the inflation.
Thus, effective values of these terms vanish if they are averaged over a time scale much longer 
than the period of the oscillation of the inflaton.
Therefore, we simply assume $a_y=a_\lambda=b=0$ after the inflation.
The fifth line is thermal-mass corrections~\cite{Allahverdi:2000zd}.
These are given by
\begin{eqnarray}
	\label{eq:thermal_mass}
	V_{\rm th}(\phi) \equiv \sum_{f_k|\phi|<T} c_kf_k^2T^2|\phi|^2 .
\end{eqnarray}
Here, $f_k$ denotes coupling constants of left-handed leptons or up-type Higgs,
and $c_k$ are determined by degrees of freedom of these particles.
The temperature of the thermal plasma $T$ before the reheating ends
is estimated as
\begin{eqnarray}
	\label{eq:temperature}
	T \sim \left(g_*^{-\frac{1}{2}}H T_R^2 M_{\mathrm{Pl}}\right)^\frac{1}{4},
\end{eqnarray}
where $g_* \simeq 200$ is the effective total degree of freedom of the thermal bath
and $M_{\rm Pl} \simeq 2.4 \times 10^{18}$\,GeV is the reduced Planck mass.
There are also thermal-log corrections \cite{Anisimov:2000wx},
but they do not have dominant effects in this leptogenesis.

During the inflation,
the potential of $\tilde{N}$ has minima at $|\tilde{N}| \simeq \sqrt{HM_{\mathrm{Pl}} / \lambda}$.
However, starting from these minima,
$\tilde{N}$ is eventually trapped at minima of $F$-term potential,
$|\tilde{N}|\simeq \sqrt{M_NM_{\mathrm{Pl}}/ \lambda}$.
In order to avoid this disastrous consequence, we assume that $\tilde{N}$ is in a multiplet of
the $SO(10)$ Grand Unified Theory (GUT).
When $|\tilde{N}| > M_{\mathrm{GUT}}$,
the steep $D$-term potential of the GUT gauge group appears.
Therefore, the initial value of $\tilde{N}$ is given by
$|\tilde{N} _{\rm ini}|=M_{\mathrm{GUT}}=10^{16}\,{\mathrm{GeV}}$.
In addition, it is required that the local maximum of the $F$-term potential 
of the radial direction of $\tilde{N}$, 
which is given by $|\tilde{N}|\simeq \sqrt{M_NM_{\mathrm{Pl}}/(3\lambda)}$,
is placed at $|\tilde{N}| > M_{\mathrm{GUT}}$.
This requirement can be rewritten by
\begin{eqnarray}
M_N/\lambda > 1.2\times10^{14}\,{\mathrm{GeV}} 
\label{eq:constraint2}.
\end{eqnarray}
Provided with these conditions, $\tilde{N}$ rolls down towards the origin and oscillates around there
after $H<M_N$.

\section{Evolution of scalar fields}

The evolution equations of the scalar fields are given by
\begin{eqnarray}
	\label{eq:evolution_N}
	\ddot{\tilde{N}} + (3H + \Gamma_N) \dot{\tilde{N}}
	+ \frac{\partial V}{\partial \tilde{N}^*} &=& 0,  \\
	\label{eq:evolution_phi}
	\ddot{\phi} + 3H\dot{\phi} + \frac{\partial V}{\partial \phi^*} &=& 0,
\end{eqnarray}
where $ \Gamma_N = {y_\nu^2} M_N/(4\pi) $ is the decay width of $\tilde N$.
The initial condition of $\tilde N$ is $|\tilde{N} _{\rm ini}|=M_{\mathrm{GUT}}$.
On the other hand, the initial phase of $\tilde{N}$ is
dependent on whether or not the Hubble parameter $H_{\mathrm{inf}}$
is much larger than the effective mass of the phase direction during inflation.
The potential is dominated by 
$A$ or $B$-term under the condition (\ref{eq:constraint2}).
If the effective mass of the phase direction is much smaller than $H_{\mathrm{inf}}$,
$\theta_{\tilde{N}}$ is randomly displaced from minima of the potential.
Otherwise, $\theta_{\tilde{N}}$ may be fixed at one of minima of the potential.
In both cases, the successful baryogenesis can be realized as we will see later.
In the former case, $H_{\mathrm{inf}}$ is constrained to be 
$H_{\mathrm{inf}} < 3\times10^{12}{\mathrm{GeV}}$
for small enough baryonic isocurvature perturbation as discussed in Section 5.2.
In the latter case, large $M_N$ or $\lambda$ is required
to derive the large effective mass of the phase direction.
Meanwhile, $\phi$ has positive Hubble-induced mass and large effective mass
$y_\nu M_{\mathrm{GUT}}$ via the neutrino Yukawa coupling.
Therefore, $\phi$ is trapped at the origin $\phi=0$ and only has quantum fluctuation
suppressed by its large effective mass.
Although the value $\langle \phi \rangle$ averaged over the universe
is expected to vanish, a finite value $\langle \phi^2 \rangle$ remains.

\subsection{Before destabilization of $\phi$}

For $H>M_N$, $\tilde N$ remains as $|\tilde N| = |\tilde N_{\rm ini}|$.
When $H$ becomes $H<M_N$, the mass term dominates the evolution of $\tilde{N}$
and $\tilde{N}$ begins oscillation around the origin.
The amplitude of the oscillation decreases with
\begin{eqnarray}
	\label{eq:evolution_N-phase1}
	|\tilde{N}| \sim M_{\mathrm{GUT}}\frac{H}{M_N}.
\end{eqnarray}

The cross term in the $F$-term potential $(y_\nu M_N \phi^2{\tilde{N}}^*  + h.c.)$
gives negative contribution to the effective mass of $\phi$.
Therefore, when $H$ decreases to $H_1$, 
$\phi$ becomes tachyonic and acquires large amplitude
because the mass squares from $H^2|\phi|^2$, $y_\nu^2 |\phi|^2|\tilde{N}|^2$ and 
the cross term decreases as $H^2$, $H^2$ and $H$, respectively.
The negative mass dominates over the Hubble induced mass when 
$H < y_\nu M_{\mathrm{GUT}}$,
while it dominates over the effective mass from the quartic coupling when
$H < {M_N^2}/ ({y_\nu M_{\mathrm{GUT}}})$.
The $LH_u$ direction $\phi$ becomes tachyonic when both conditions are satisfied.
Therefore, we define the Hubble parameter at the destabilization $H_1$ as
\begin{eqnarray}
	\label{eq:destabilization}
	H_1 = 
	\left\{
	\begin{array}{ll}
	y_\nu M_{\mathrm{GUT}} & ~~ 
	\left(m_\nu < 10^{-8} \, {\rm eV}
	\left(\frac{M_N }{10^{11} \, \rm GeV } \right) \right) \\
	\\
	\displaystyle\frac{M_N^2}{y_\nu M_{\rm GUT}} &~~ 
	\left(m_\nu > 10^{-8} \, {\rm eV}
	\left(\frac{M_N }{10^{11} \, \rm GeV } \right) \right)
	\end{array}
	\right. .
\end{eqnarray}
Note that always $H_1<M_N$.
The negative contribution from $\phi^2\tilde{N}^{*3}$ term
cannot be dominant under the condition (\ref{eq:constraint2}).

Thermal-mass terms give positive contributions to the effective mass of $\phi$,
while thermal-log corrections are ineffective because $\phi$ does not have large value.
The destabilization of $\phi$ is prevented if thermal-mass corrections
dominate over the negative mass contribution.
Otherwise, the same scenario as discussed 
in Ref.\,\cite{Allahverdi:2004ix} is realized.
This condition is given by
\begin{eqnarray}
	\label{eq:thermal-constraint}
	H_1 > H'_1 \simeq 10^7 \,{\mathrm{GeV}}
	\left( \frac{ g_* }{ 200 }\right)^{-\frac{1}{2}}
	\left( \frac{T_R}{10^9{\mathrm{GeV}}} \right)^2
	\left( \frac{y_\nu}{10^{-2}} \right)^{-2}
	\left( \frac{M_{\mathrm{GUT}}}{10^{16}{\mathrm{GeV}}} \right)^{-2},
\end{eqnarray}
where we consider the thermal-mass correction from top (s)quarks in the thermal bath.

In addition, the right-handed sneutrino must not decay before the destabilization.
This constraint is given as $\Gamma_N < H_1$.
If $H_1 = y_\nu M_{\mathrm{GUT}}$, 
this is rewritten by $ M_N < 10^{16} {\rm \, GeV} (m_\nu /\rm 10^{-2} \, eV)^{-1/3}$.
On the other hand, for $H_1 = M_N^2/(y_\nu M_{\mathrm{GUT}})$,
this constraint is given as $ M_N < 10^{16} {\rm \, GeV} (m_\nu /\rm 10^{-2} \, eV)^{-3}$,
Thus, this constraint is ignorable in both cases.

In Ref.\,\cite{Allahverdi:2004ix}, it is also discussed that 
the parametric resonance~\cite{parametric,Kofman:1997y}
may cause fast decay of $|\tilde{N}|$.
Since $\tilde{N}$ oscillates with large amplitude, 
it gives large oscillating effective mass of $\phi$,
unless $\tilde{N}$ has a circle-like trajectory in complex plain
due to large angular momentum.
This may result in exponential amplification of fluctuation of $\phi$
and rapid decrease of the amplitude of the $\tilde{N}$ oscillation.
If $\tilde{N}$ decays via the parametric resonance before the destabilization of $\phi$,
the destabilization cannot take place.
However, this is not serious for $m_\nu > 10^{-8} \, {\rm eV} (M_N / 10^{11} \, \rm GeV)$,
because $\phi$ can decay into other particles.
The typical amplification rate under broad parametric resonance is estimated to be
proportional to $\exp(0.175M_Nt)$~\cite{Kofman:1997y}.
On the other hand, the decay rate $\Gamma_\phi$ has the smallest value
just before the destabilization of $\phi$, $\Gamma_\phi \sim \sum g_i^2 M_N/(8\pi)$,
where $g_i$ are Yukawa and gauge couplings, and we sum over all final states.
Hence, the parametric resonance is safely negligible
since $\Gamma_\phi > 0.175 M_N$ in the MSSM.
On the other hand, parametric resonance may take place
for $ m_\nu < 10^{-8} \, {\rm eV} ( M_N / 10^{11} \, \rm GeV)$,
because $\Gamma_\phi >0.175M_N$ is not guaranteed.
Since the detail of the parametric resonance is involved,
we do not discuss whether the parametric resonance
actually gives fast decay of $|\tilde{N}|$.
For simplicity, we neglect the case
$ m_\nu < 10^{-8} \,{\rm eV} ( M_N / 10^{11} \, \rm GeV)$.

\subsection{After destabilization of $\phi$}

After $H=H_1$, the potential of $\phi$ has two distinct minima
in opposite phase directions,
\begin{eqnarray}
	|\phi|\sim\sqrt{\frac{2M_N |\tilde{N}(H_1)|}{y_\nu}}
	\sim\sqrt{\frac{2M_{\mathrm{GUT}}H_1}{y_\nu}}.
\end{eqnarray}
The initial condition at the destabilization is determined by
quantum fluctuation during the inflation and the subsequent evolution.
After the destabilization of $\phi$, 
long wavelength modes of fluctuation of $\phi$ get
large value via tachyonic instability \cite{tachyonic,GarciaBellido:2002aj}.
Once the value of $\phi$ begins to track the minimum of the potential,
it can be interpreted as a classical field in a local patch of the universe.
Hereafter we assume that the universe consists of patches
in which condensate of $\phi$ has various initial values.
Hence, the resultant lepton asymmetry is estimated
by averaging over results from various values of initial quantum fluctuation.
Since a typical comoving wavelength of this quantum fluctuation is
negligibly small compared with the present horizon scale,
this fluctuation has no cosmologically observable consequence.
In other words, the initial fluctuation of $\phi$ averaged over cosmological scale is suppressed to be
negligibly small, because such scale is far larger than the horizon scale at the epoch $H=H_1$.
Indeed, the comoving length $k_1^{-1}$ of the latter scale is estimated to be
\begin{eqnarray}
	\label{eq:scale}
	k_1^{-1}\sim {\mathcal{O}}(10)\,{\mathrm{km}}\times
	\left( \frac{H_1}{10^8{\mathrm{GeV}}} \right)^{-\frac{1}{3}}
	\left( \frac{T_R}{2\times10^6{\mathrm{GeV}}} \right)^{-\frac{1}{3}}
	\left( \frac{g_*}{100} \right)^{-\frac{1}{12}}.
\end{eqnarray}	
Note that the problem of domain walls separating two minima is not serious
since they disappear when $\phi$ evaporates.

With the minimization of the cross term, 
the dominant contributions in this epoch can be rewritten by
\begin{eqnarray}
	\label{eq:F-term}
	V_F(\phi,\tilde{N}) = 
	\left( \frac{y_\nu}{2} |\phi|^2 - M_N |\tilde{N}| \right)^2
	+ y_\nu^2 |\phi|^2|\tilde{N}|^2.
\end{eqnarray}
Though this potential has the global minimum at $|\phi|=|\tilde{N}|=0$,
there is a valley on the trajectory $y_\nu|\phi|^2 = 2M_N|\tilde{N}|$,
lifted by $y_\nu^2 |\phi|^2|\tilde{N}|^2$.
Therefore, $\phi$ and $\tilde{N}$ oscillate around the origin approximately satisfying 
the relation, $y_\nu|\phi|^2 = 2M_N|\tilde{N}|$.
The amplitudes of $|\phi|$ and $|\tilde{N}|$ decrease proportionally to
$H^{1/2}$ and $H$, respectively.

The condensate of $\tilde{N}$ decays by the neutrino Yukawa coupling at $H \sim \Gamma_N$.
Since the decay rate acts as a friction term in the evolution equation,
$\tilde{N}$ is strongly fixed at the minimum of the potential
soon after $H = \Gamma_N$.
Hence, the evolution of the scalar fields is determined by that of $\phi$.
After $| \tilde N |$ is integrated out by $|\tilde{N}| = y_\nu |\phi|^2 / (2 M_N)$,
the effective potential of $\phi$ is given by
\begin{eqnarray}
	\label{eq:effective_potential}
	V_{\mathrm{eff}} (\phi) \simeq \frac{y_\nu^4}{4}\frac{|\phi|^6}{M_N^2}
		+ V_{\rm th} (\phi),
\end{eqnarray}
where we included thermal potentials.
The $LH_u$ direction does not decay before $\tilde{N}$ does,
since all decay modes with coupling constants larger than $y_\nu$
are kinematically forbidden as long as $H > \Gamma_N$.
Since the first term decreases as $H^6$,
thermal-corrections usually dominate the evolution of $\phi$,
soon after the $\tilde{N}$ decay starts.
Afterward, $\phi$ oscillates around the origin and eventually evaporates.

\begin{figure}[t]
	\begin{center}			
		\includegraphics[width=.9\linewidth]{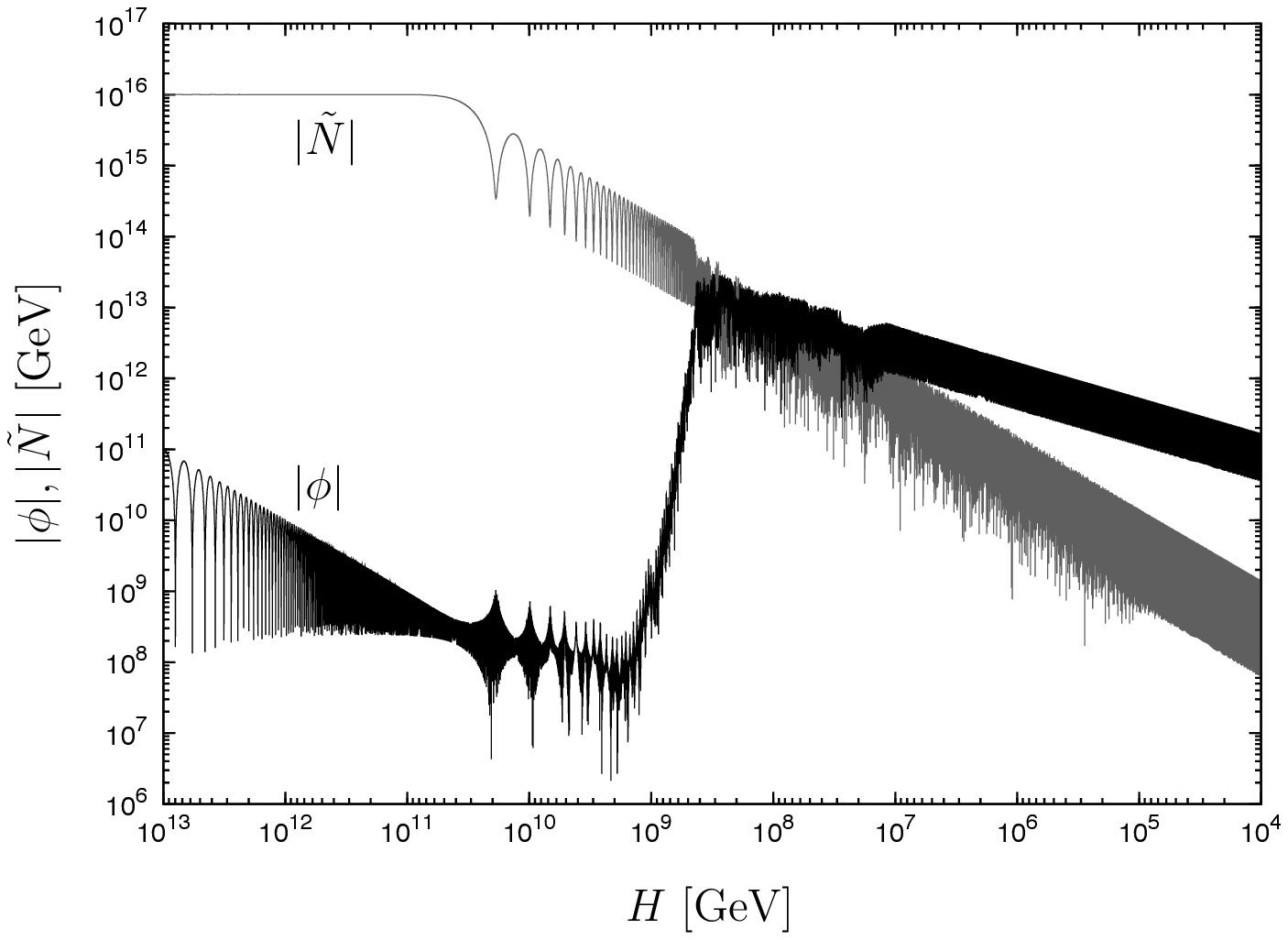}
		\caption{\small The evolution of the values of
		$|\phi|$ (black) and $|\tilde{N}|$ (grey)
		as a function of $H$.
		We take $M_N=10^{11} \, {\mathrm{GeV}}$, $y_\nu=10^{-2}$,
		$T_R=2\times10^6 \, {\mathrm{GeV}}$ and $\lambda = 10^{-4}$.
		}
	\label{fig:evolution-F}
	\end{center}
\end{figure}

In Fig.\,\ref{fig:evolution-F}, we show an example of the evolution of the scalar fields.
The black (grey) line show the evolution of $|\phi| (|\tilde{N}|)$.
Here, parameters are taken as $M_N=10^{11} \, {\mathrm{GeV}}$, $y_\nu=10^{-2}$,
$T_R=2\times10^6 \, {\mathrm{GeV}}$, $c_\phi = c_N = 1$ and $\lambda = 10^{-4}$.
We assumed $a_y = a_\lambda =b=0$ as mentioned above.
As the initial condition, we assumed
$ | \tilde N _{\rm ini} | = M_{\rm GUT} ~(\arg \tilde{N}_{\mathrm{ini}} = \pi / 4)$ and 
$|\phi_{\mathrm{ini}}|=10^{11} \, {\mathrm{GeV}}$.
Here we evaluated typical value of quantum fluctuation 
by averaging over the horizon scale after the inflation
with $H_{\mathrm{inf}} = 10^{13}\,{\mathrm{GeV}}$,
because the destabilization of $\phi$ is not instantaneous.
In order to take the randomness of the initial condition into account, 
we iterated the same calculation for various initial amplitudes and phases of $\phi$.
This figure is the result for ${\mathrm{arg}}(\phi_{\mathrm{ini}})=0$.
It can be seen that $|\tilde{N}|$ is fixed
at $|\tilde{N}|=M_{\mathrm{GUT}}$ until $H\simeq M_N/3$ and 
begins oscillation at $H\simeq M_N/3$ from this figure.
We can also confirm that
the destabilization of $\phi$ completes at $H\sim H_1 =10^{8} \, {\mathrm{GeV}}$.
Afterward, $|\phi|$ decreases proportionally to $H^{1/2} $,
while $|\tilde{N}|$ does proportionally to $H$.
Finally, $\tilde{N}$ decays at $H\sim\Gamma_N$.

\section{Evolution of lepton asymmetry}

The evolution of the lepton asymmetry is determined by the evolution of the scalar fields.
We define lepton asymmetries
in the condensate of the right-handed sneutrino $L_{\tilde{N}}$
and in that of the $LH_u$ direction $L_\phi$ as follows,
\begin{eqnarray}
	L_{\tilde{N}} &\equiv& -i(\dot{\tilde{N}}^*\tilde{N} - \dot{\tilde{N}}\tilde{N}^*), \\
	L_\phi &\equiv& \frac{i}{2}(\dot{\phi}^*\phi - \dot{\phi}\phi^*).
\end{eqnarray}	
Note that we assigned lepton number $-1$ to $\tilde N$. 
Evolution equations of these lepton asymmetries are given from the evolution equations
of these scalar fields, Eqs.\,(\ref{eq:evolution_N}) and (\ref{eq:evolution_phi}),
\begin{eqnarray}
	\label{eq:evolution_nN}
	&&\frac{d}{dt}\left( \frac{L_{\tilde{N}}}{H^2} \right) 
	+\Gamma_N\frac{L_{\tilde{N}}}{H^2} \nonumber \\
	&& ~~~~~~~~~~~~~= - \frac{1}{H^2}\left[ y_\nu M_N {\mathrm{Im}} ({\phi^*}^2\tilde{N}) 
	+ \frac{4\lambda M_N}{M_{\mathrm{Pl}}}{\mathrm{Im}}\left( \tilde{N}^3\tilde{N}^* \right)\right.
	\nonumber \\
	&& ~~~~~~~~ ~~~~~~~~ ~~~~~~~~ ~~~~~~~~ ~~~~~~~~ ~~~~~~~~
	+ \left.\frac{3y_\nu \lambda}{M_{\mathrm{Pl}}}{\mathrm{Im}}\left( \phi^{*2} \tilde{N}^3 \right)
	\right], \\
	&&~\nonumber\\
	\label{eq:evolution_nL}
	&&\frac{d}{dt}\left( \frac{L_\phi}{H^2} \right) =
	- \frac{1}{H^2} \left[ y_\nu M_N {\mathrm{Im}} ({\phi^*}^2\tilde{N})
	+ \frac{y_\nu \lambda}{M_{\mathrm{Pl}}}{\mathrm{Im}}\left( \phi^{*2} \tilde{N}^3 \right) \right] ,
\end{eqnarray}	
where Hubble-induced phase-dependent terms are dropped as mentioned above.
The right-hand side of these equations are source terms of the lepton asymmetry.
If a dominant source term is simply scaling with
$t^\gamma \propto H^{-\gamma}$ on an average, 
where the index $\gamma$ is a constant,
the evolution of the lepton asymmetry is simple:
if $\gamma>-3$, the lepton asymmetry increases proportionally to $t^{\gamma+3}$,
while if $\gamma<-3$, the lepton asymmetry is fixed.
In the case of $\gamma=-3$, the lepton asymmetry increases proportionally to $\log t$.
Thus, the lepton asymmetry is also considered to be almost fixed in this case.

The evolution of these lepton asymmetries are very complicated.
However, the evolution of the left-right asymmetry $L_\phi - L_{\tilde{N}}$ is simpler.
The evolution of $L_\phi - L_{\tilde{N}}$ is given by the equation,
\begin{eqnarray}
	\label{eq:evolution_LR}
	\frac{d}{dt}\left( \frac{L_\phi - L_{\tilde{N}}}{H^2} \right) =
	\frac{1}{H^2}\left[ 
	\frac{4\lambda M_N}{M_{\mathrm{Pl}}}{\mathrm{Im}}\left( \tilde{N}^3\tilde{N}^* \right)
	+ \frac{2y_\nu \lambda}{M_{\mathrm{Pl}}}{\mathrm{Im}}\left( \phi^{*2} \tilde{N}^3 \right) \right] 
	~~~~\nonumber \\
	+ \Gamma_N \frac{L_{\tilde{N}}}{H^2}.
\end{eqnarray}	
Thus, the dominant contribution from $\phi^{*2}\tilde{N}$ term to the left-right asymmetry cancels.
Since the other terms attenuates sufficiently fast,
the evolution of the left-right asymmetry is fixed after $\tilde{N}$ begins oscillation.

\subsection{Before destabilization of $\phi$}

In this era, the decay width is negligible because $H \gg \Gamma_N$.
The second and third terms in the right-hand side of Eq.\,(\ref{eq:evolution_LR}) are also negligible,
since the fluctuation of $\phi$ is suppressed by its large effective mass.
Thus, the first term is dominant because the phase direction of $\tilde{N}$
is generally displaced from minima determined by $\tilde{N}^3\tilde{N}^*$ term.
Therefore, $L_\phi - L_{\tilde{N}}$ is determined only by the evolution of $\tilde{N}$.
The magnitude of left-right asymmetry $L_\phi - L_{\tilde{N}}$ becomes fixed 
when $\tilde{N}$ begins oscillation,
because the amplitude of the source term scales with $H^4$.
The amount of the left-right asymmetry at $H=M_N / 3$ is estimated by
\begin{eqnarray}
	\label{eq:left-right}
	\frac{ | L_\phi - L_{\tilde{N}} | }{s'} \simeq \frac{ | L_{\tilde{N}} | }{s'}
	\sim \frac{6\lambda M_{\mathrm{GUT}}^4  T_R}{M_N^2M^3_{\mathrm{Pl}}}\delta_{\mathrm{eff}},
\end{eqnarray}	
where the entropy parameter $s'$ is defined as $s'\equiv 4M_{\mathrm{Pl}}^2 H^2/T_R$,
and $\delta_{\mathrm{eff}} \lesssim 1$ is the phase factor.
Note that $s'$ is normalized by the entropy after the reheating completes.

%

\subsection{After destabilization of $\phi$}

After $\phi$ acquires large value, it takes part in the evolution.
In Eqs.\,(\ref{eq:evolution_nN}) and (\ref{eq:evolution_nL}),
the first term of the right-hand side becomes dominant.
Since this term induces rapid exchange between $L_{\tilde{N}}$ and $L_\phi$,
these asymmetries oscillate rapidly.

However, $L_\phi - L_{\tilde{N}}$ is fixed
because all the source terms for left-right asymmetry attenuate with $H^4$
after $\phi$ gets large value.
At the epoch $\phi$ is rolling down to the displaced minimum, 
the contribution to the $L_\phi - L_{\tilde{N}}$
from the second term of Eq.\,(\ref{eq:evolution_LR}) is difficult to estimate 
because of the fast and non-linear evolution.
Furthermore, this contribution depends on the initial phase of random fluctuation of  $\phi$.
However, we confirmed that this contribution is subdominant
if the contribution from the first term is sufficiently large,
since the contribution from the second term almost vanishes
on an average over cosmological scale,
which contains large number of patches of horizon scale at the epoch $H=H_1$.
In other words, the initial fluctuation of $\phi$ averaged over cosmogical scale is suppressed to be
negligibly small, because such scale is far larger than the horizon scale at the epoch $H=H_1$.
After $H \sim \Gamma_N$, $\tilde{N}$ decays rapidly by the neutrino Yukawa coupling.
The leading decay channels of $\tilde{N}$ are
$\tilde{N} \to H_u \tilde{L}$ and $\tilde{N} \to \tilde{H}_u \bar{L}$,
creating lepton numbers $+1$ and $-1$, respectively.
Therefore, the lepton asymmetry stored in the condensate of $\tilde{N}$
is not transfered to the SM sector if $CP$-violation in the decay is small.
On the other hand,
$L_\phi$ is fixed because the source term decreases rapidly after $H \sim \Gamma_N$.
The amount of the fixed asymmetry of $L_{\phi}$ is
the same order of $ L_\phi - L_{\tilde{N}} $.
The precise amount is dependent on the phase of the oscillation of $L_{\phi}$
and difficult to estimate analytically.
Therefore, We define $\epsilon $
as the fraction of $ L_\phi - L_{\tilde{N}} $ inherited to $L_\phi$.
The fixed asymmetry of $L_{\phi}$ also depends on the initial value of $\phi$.
Since the value of $ L_\phi - L_{\tilde{N}} $
is determined only by the evolution of ${\tilde{N}}$,
non-vanishing asymmetry is inherited in $L_\phi$
after averaging results for various initial values of $\phi$.
Eventually, $L_\phi$ is released to the SM sector
as the condensate $\phi$ evaporates.

The fact that the final lepton asymmetry have non-vanishing value can also be understood
qualitatively by the following discussion.
The potential of $\phi$ has two distinct minima after destabilization due to coupling with $\tilde{N}$.
The crucial observation is that the direction of the rotational evolution of $\tilde{N}$
is determined only by initial evolution of $\tilde{N}$,
therefore this direction is the same over the cosmological scale.
Since $\tilde{N}$ rotates in a definite direction, these two minima also rotate in one direction.
Because the rotational motion of $\phi$ is induced by this rotation of two minima,
$\phi$ rotates in one direction whichever is the minimum that $\phi$ is trapped.
After $\tilde{N}$ decays, the direction of the rotation of $\phi$ does not change 
because of approximate conservation of the angular momentum.
Thus, the final direction of the rotation of $\phi$, 
which is equivalent to the sign of the final lepton asymmetry $L_\phi$,
is the same all over the universe.

\begin{figure}[t]
	\begin{center}			
		\includegraphics[width=.9 \linewidth]{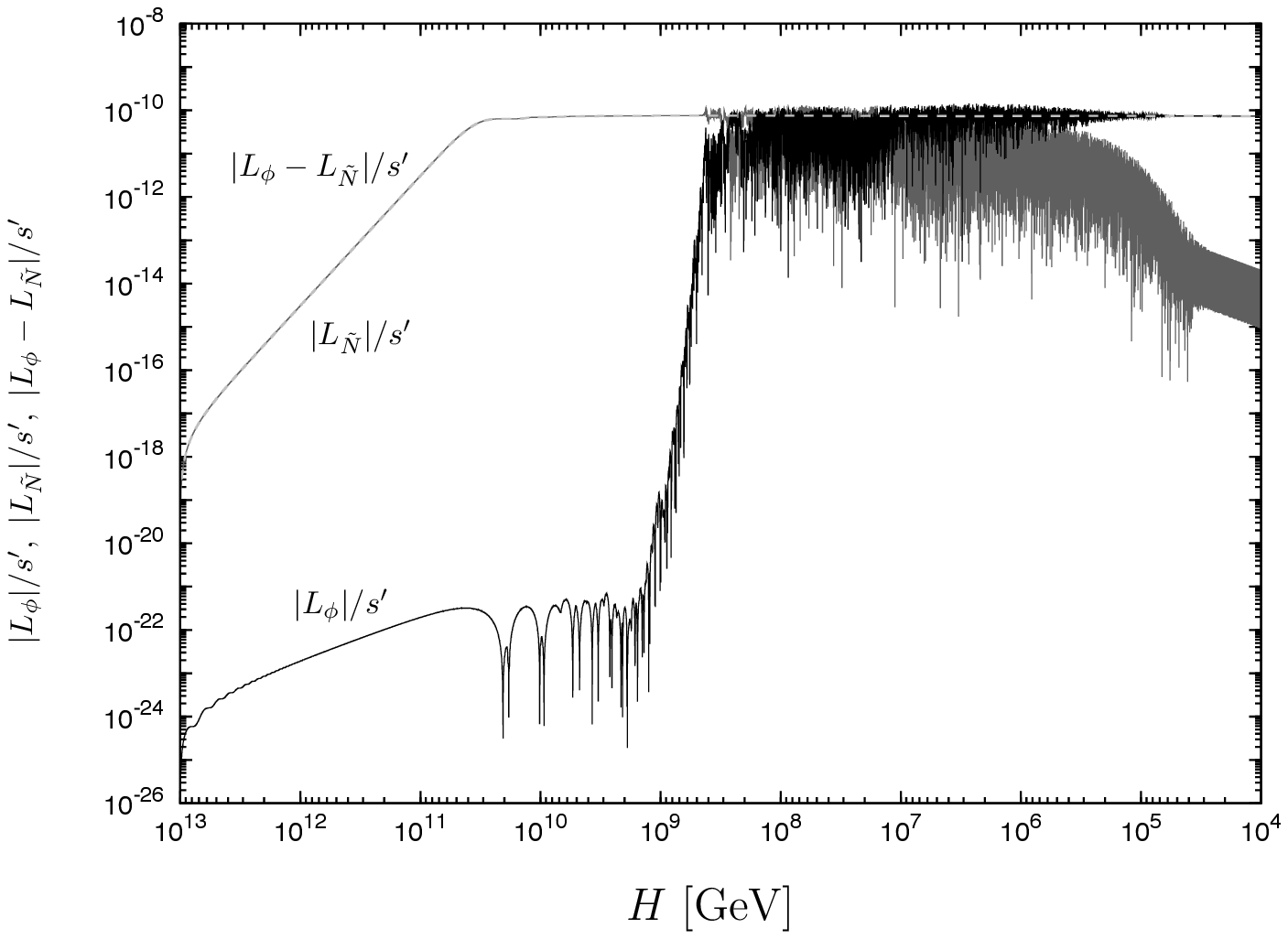}
		\caption{\small The evolution of lepton asymmetries $|L_\phi|/s'$ (solid black),
		$|L_{\tilde{N}}|/s'$ (solid grey) and 
		 the left-right asymmetry $|L_\phi-L_{\tilde{N}}|/s'$ (dashed grey).
		 Since these asymmetries oscillate, we show magnitudes of them.
		 The parameters are the same as for Fig.\,\ref{fig:evolution-F}.
		}
	\label{fig:evolution-L}
	\end{center}
\end{figure}

We show the evolution of asymmetries in Fig.\,\ref{fig:evolution-L}
for the same parameter set as we chose for Fig.\,\ref{fig:evolution-F}.
The solid black line, the solid grey line, and the dashed grey line indicate evolutions of
$|L_\phi|/s'$, $|L_{\tilde{N}}|/s'$ and $|L_\phi-L_{\tilde{N}}|/s'$, respectively.
It can be seen that for $H > M_N$,
$|L_{\tilde N}|$ increases, while $|L_\phi|$ is negligible.
At $H \sim H_1$, $\phi $ receives a fraction of the asymmetry of $\tilde N$.
We can also confirm from this figure that for $H < \Gamma_N \simeq 10^6 $\,GeV,
$L_{\tilde N}$ decreases exponentially and $L_\phi$ is fixed.
For the parameter used in our calculation,
the estimation using Eq.\,(\ref{eq:left-right}) gives
$|L_\phi |/s = 8.3 \times 10^{-11} \delta_{\rm eff} \epsilon$.
We iterated the same calculation for 25 various initial phases,
and the result is $| L_\phi |/s = 5.2\times10^{-11}$ on an average.
This implies $\delta_{\mathrm{eff}}\epsilon \sim0.6$ in this case.
We also confirmed that other choices of initial amplitude $|\phi_{\mathrm{ini}}|$
do not change the result.

\section{Resultant baryon asymmetry}

\subsection{Estimation of baryon asymmetry}

The lepton asymmetry released into the SM sector is transfered to the baryon asymmetry
through the sphaleron process by the ratio $n_B = (8/23) n_L$ \cite{Harvey:1990qw},
where $n_B$ is the resultant baryon asymmetry in the thermal bath
and $n_L$ is the lepton asymmetry produced by this scenario.
Hence, the baryon asymmetry is estimated as
\begin{eqnarray}
	\label{eq:baryon_asymmetry}
	\frac{n_B}{s'} \sim \frac{48}{23}\frac{\lambda M_{\rm GUT}^4 T_R}{M_N^2 M^3_{\mathrm{Pl}}}
	\delta_{\mathrm{eff}}\epsilon .
\end{eqnarray}	

\begin{figure}[t]
	\begin{center}			
		\includegraphics[width=.9\linewidth]{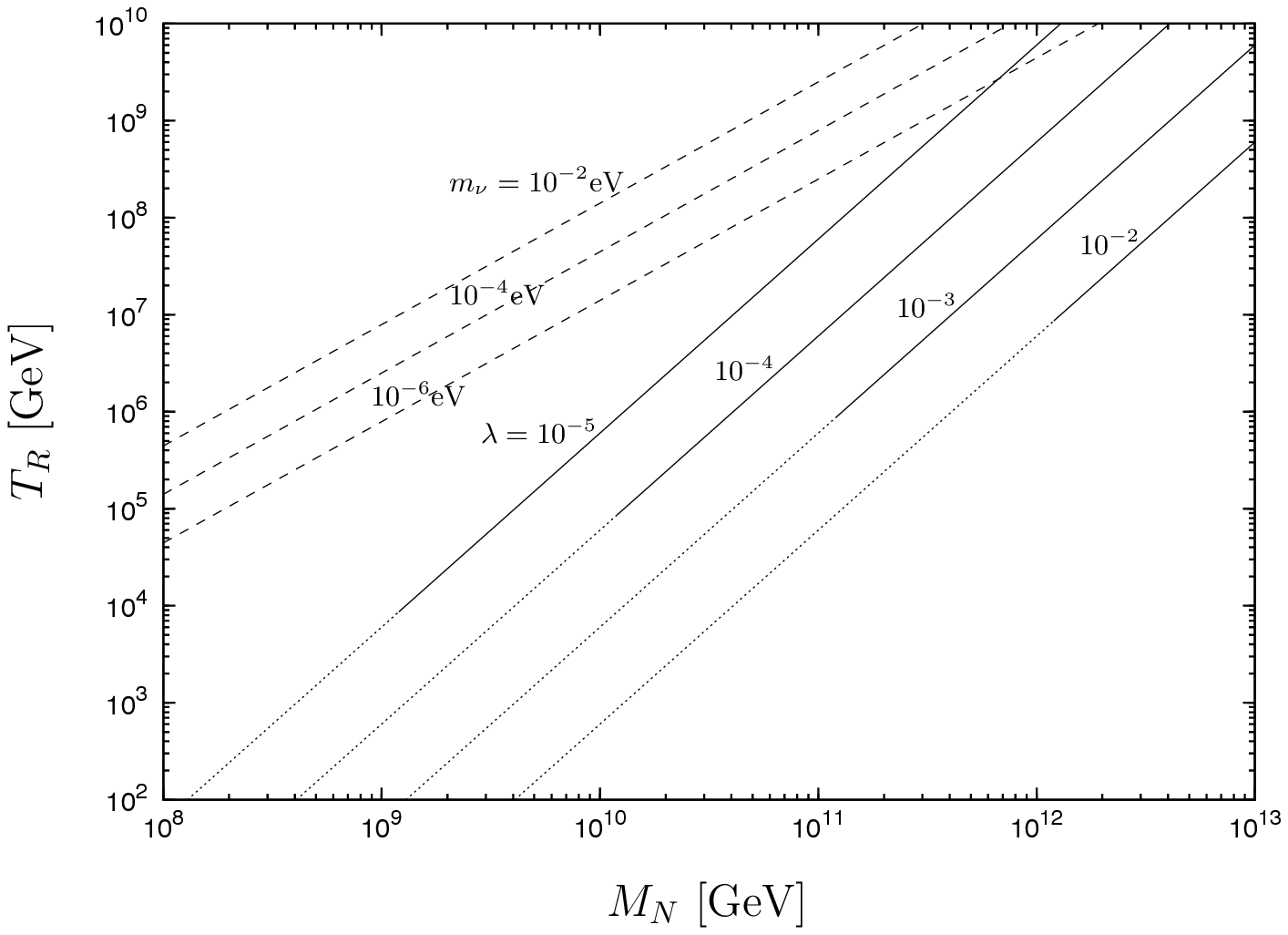}
		\caption{\small Solid lines show parameter region in which
		the rusultant baryon asymmetry is $n_B/s=8.7\times10^{-11}$
		for various values of $\lambda$.
		We assumed that $\delta_{\mathrm{eff}}\epsilon=1$.
		Dotted parts of these lines are excluded by the condition $M_N/\lambda > 1.2\times10^{14}
		\,{\mathrm{GeV}}$.
		Dashed lines show the constraint $H>H'_1$ for several $m_\nu$,
		which is not stringent.
		Above these lines, the scenario discussed in Ref.\,\cite{Allahverdi:2004ix}
		can be realized.
		}
	\label{fig:contour1}
	\end{center}
\end{figure}

In Fig.\,\ref{fig:contour1}, we show the parameter region in which
the right amount of baryon asymmetry is produced
and some constraints on the $M_N$-$T_R$ plane.
Solid lines with dotted parts show region in which resultant baryon asymmetry 
(\ref{eq:baryon_asymmetry}) is $n_B/s=8.7\times10^{-11}$ for various $\lambda$.
We assumed that $\delta_{\mathrm{eff}}\epsilon=1$.
The baryon asymmetry is simply proportional to this factor.
Dotted parts of these lines are excluded because of the constraint (\ref{eq:constraint2}).
Dashed lines indicate the constraint that thermal-corrections must not 
dominate over the negative mass contribution of $\phi$ for several neutrino masses
(see Eqs.\,(\ref{eq:destabilization}) and (\ref{eq:thermal-constraint})).
Note that neutrino masses are determined by Eq.\,(\ref{eq:seesaw}).
We can see that this constraint is not stringent.
Above these lines, the scenario discussed in Ref.\,\cite{Allahverdi:2004ix}
can be realized.
In addition, if initial $\theta_{\tilde{N}}$ is not fixed at 
any minima determined by Hubble-induced $A$- or $B$-term,
$M_N<H_{\mathrm{inf}}<3\times10^{12}{\mathrm{GeV}}$ is required
in order that baryonic isocurvature perturbation should be sufficiently small (see Section 5.2).
This result indicates that successful baryogenesis via this scenario
favors larger $M_N$ and higher $T_R$ for large $\lambda$
and smaller $M_N$ and lower $T_R$ for small $\lambda$.

As we can see in Eq.\,(\ref{eq:baryon_asymmetry}),
the resultant baryon asymmetry is proportional to
$|\tilde{N}_{\mathrm{ini}}|^4 = M_{\rm GUT}^4$.
If $\tilde N$ is the multiplet of the subgroup of SO(10) GUT, e.g. SU(2$)_R$,
$|\tilde{N}_{\mathrm{ini}}|=M_{SU(2)_R}<M_{\mathrm{GUT}}$ should be used.
Unless $\lambda$ is extremely small, it is difficult to explain the origin of baryon asymmetry
by this scenario for $|\tilde{N}_{\mathrm{ini}}| = M_{SU(2)_R} < M_{\mathrm{GUT}}$.

Finally, we summarize the difference
between our scenario and that in Ref.\,\cite{Allahverdi:2004ix}.
The latter scenario may be realized above dashed lines in Fig.\,\ref{fig:contour1}.
Our scenario can explain the baryon asymmetry for lower $T_R$
than that in Ref.\,\cite{Allahverdi:2004ix}.
On the other hand, the scenario in Ref.\,\cite{Allahverdi:2004ix}
has an advantage that the sufficient baryon asymmetry can be generated
if $|\tilde{N}_{\mathrm{ini}}|=M_{SU(2)_R}<M_{\mathrm{GUT}}$.
Another advantage of their scenario is that non-renormalizable terms in the superpotential
is not required.
On the other hand, the most important advantage of our scenario is that 
the successful baryogenesis can be realized without fine-tuning.
The scenario in Ref.\,\cite{Allahverdi:2004ix} requires
some tuning between the decay width of $\Gamma_N$ and the magnitude of $B$-terms.
However, any parameter tuning is not required in our scenario.

\subsection{Baryonic isocurvature perturbation}
In the case that the potential of the phase direction $\theta_{\tilde{N}}$ of $\tilde{N}$
is sufficiently flat during the inflation,
$\theta_{\tilde{N}}$ is randomly displaced from minima of the potential.
Then, $\theta_{\tilde{N}}$ has isocurvature perturbation,
\begin{eqnarray}
	\label{eq:isocurvature}
	\delta\theta_{\tilde{N}} = \frac{H_{\mathrm{inf}}}{\sqrt{2k^3}|\tilde{N}|}.
\end{eqnarray}	
for Fourier mode $k$.
Since $L_{\tilde{N}}$ is produced via the displacement of $\theta_{\tilde{N}}$ from 
minima determined by $\tilde{N}^3\tilde{N}^*$ term,
this isocurvature perturbation results in the isocurvature perturbation of $L_{\tilde{N}}$,
which is finally transfered to the isocurvature perturbation of baryon asymmetry.
Since $n_B\propto \delta_{\mathrm{eff}}$ and $\delta_{\mathrm{eff}}$ can be estimated by
$\delta_{\mathrm{eff}}\sim \sin(2\delta\theta_{\tilde{N}})$,
the amplitude of the baryonic isocurvature perturbation can be estimated to be
\begin{eqnarray}
	\label{eq:baryonic_isocurvature}
	\frac{\delta n_B}{n_B}\sim 2\delta\theta_{\tilde{N}}.
\end{eqnarray}	
According to the constraint on the baryonic isocurvature perturbation 
in terms of the ratio between the power spectrum of matter isocurvature perturbation 
and that of curvature perturbation \cite{Kawasaki:2007mb},
\begin{eqnarray}
	\label{eq:consraint_baryonic_isocurvature}
	B_a \equiv \sqrt{\frac{\mathcal{P_S}}{\mathcal{P_R}}}
	= \sqrt{\frac{1}{2.4\times10^{-9}}}\sqrt{\frac{k^3}{2\pi^2}
	\frac{\Omega_b^2}{\Omega_m^2}\left\langle \frac{\delta n_B^2}{n_B^2} \right\rangle}
	<0.31,
\end{eqnarray}	
the Hubble paraneter during the inflation is constrained to be 
$H_{\mathrm{inf}}<3\times10^{12}\,{\mathrm{GeV}}$.
Note that this constraint can be avoided
if initial $\theta_{\tilde{N}}$ is fixed at one of minima determined by Hubble-induced $A$- or $B$-term.
This can be realized if the value of $ b $
or $a_\lambda$ during the inflation satisfies the following condition,
$bM_N \sim H_{\mathrm{inf}}$ or $\lambda a_\lambda > (H_{\mathrm{inf}}/10^{14}\,{\mathrm{GeV}})$.

\section{Summary}

We reconsidered the leptogenesis scenario in the SUSY seesaw model
with including the evolution of the $LH_u$ direction.
We found that the $LH_u$ direction acquires large value
due to a negative effective mass induced by the right-handed sneutrino condensate
through the Yukawa coupling, 
even if the minimum of $\phi$ is fixed at the origin during the inflation,
unless the reheating temperature $T_R$ is sufficiently high.
The lepton asymmetry is first produced in the condensate of $\tilde{N}$
via the Affleck-Dine mechanism, then
transfered nonperturbatively to the condensate of $\phi$ by the Yukawa coupling.
This lepton asymmetry is released into the baryon asymmetry in the SM sector
by the sphaleron process.
In this scenario, the appropriate amount of the baryon asymmetry can be generated
for low $T_R$ avoiding the gravitino problem without parameter tuning.

\section*{Acknowledgments}

The work of MS is supported in part by the Grant-in-Aid
for Science Research, Ministry of Education, Science and Culture,
Japan (No.~18840011).


\end{document}